\documentclass[twocolumn,5p,times,sort&compress]{elsarticle}
\usepackage[T1]{fontenc}
\usepackage[utf8]{inputenc}

\usepackage{amsmath,amsfonts,amssymb,bm}
\usepackage[colorlinks,linkcolor=blue,citecolor=blue,urlcolor=blue]{hyperref}

\usepackage[english]{babel}
\newcommand{\qbar}{{\bar q}}

\newcommand{\Pqx}{P^q_x}
\newcommand{\Pqy}{P^q_y}
\newcommand{\Pqz}{P^q_z}

\newcommand{\Pqbarx}{P^\qbar_x}
\newcommand{\Pqbary}{P^\qbar_y}
\newcommand{\Pqbarz}{P^\qbar_z}

\newcommand{\ePP}{3+\mathbf{P}^q\cdot\mathbf{P}^\qbar}  % Common factor in denominators.

\newcommand{\re}{\mathrm{Re}}
\newcommand{\im}{\mathrm{Im}}

\newcommand{\tr}{\mathrm{tr}}

\newcommand{\sNN}{\sqrt{s_\text{NN}}}

\newcommand{\PG}{P_\text{global}}

\newcommand{\bigO}{\text{\usefont{OMS}{cmsy}{m}{n}O}}

\begin{document}

\title{Local spin alignment of vector mesons in relativistic heavy-ion collisions}

\author[PhyDep]{Xiao-Liang Xia}
\ead{xiaxl@fudan.edu.cn}

\author[KeyLab]{Hui Li}
\ead{lihui\_fd@fudan.edu.cn}

\author[PhyDep,KeyLab]{Xu-Guang Huang}
\ead{huangxuguang@fudan.edu.cn}

\author[KeyLab,UCLA]{Huan Zhong Huang}
\ead{huanzhonghuang@fudan.edu.cn}

\address[PhyDep]{Department of Physics and Center for Field Theory and Particle Physics, Fudan University, Shanghai 200433, China}
\address[KeyLab]{Key Laboratory of Nuclear Physics and Ion-beam Application (MOE), Fudan University, Shanghai 200433, China}
\address[UCLA]{Department of Physics and Astronomy, University of California, Los Angeles, CA 90095, USA}

\begin{abstract}
We investigate the spin alignment of vector mesons arising from locally polarized quarks and anti-quarks (local spin alignment) in heavy-ion collisions. We find that $\rho_{00}\neq 1/3$ does not necessarily signal the global polarization of quarks and anti-quarks along the orbital angular momentum of the system, but may also originate from their local spin polarization. Such local spin polarization could be induced by local vorticity arising from the anisotropic expansion of the fireball. We further propose experimental observables that can distinguish between the local and the global spin alignments. Measurements of these observables in heavy-ion collisions can probe the vorticity pattern and shed light on the puzzles of $\phi$ and $K^{*0}$ spin alignments.
\end{abstract}

\maketitle

\section{Introduction}

In noncentral relativistic heavy-ion collisions, when two nuclei collide at a finite impact parameter, a large orbital angular momentum (OAM) of the order of $10^5-10^7\hbar$ can be generated~\cite{Becattini:2007sr,Deng:2016gyh,Jiang:2016woz}. It has been proposed~\cite{Liang:2004ph,Voloshin:2004ha,Betz:2007kg,Gao:2007bc,Huang:2011ru} that such an OAM can be partially transferred to the spin of quarks and anti-quarks in the produced quark-gluon plasma (QGP) due to spin-orbit coupling. Statistical mechanics and kinetic theory further show that the OAM can manifest itself in the form of fluid vorticity and polarize the particles in the system~\cite{Becattini:2013fla,Becattini:2013vja,Fang:2016vpj,Liu:2020flb}. As a result, hadrons emitted from the QGP will have a net spin polarization along the OAM direction. This phenomenon is referred to as the {\it global polarization}. Recently, the global polarization of $\Lambda$ hyperon in Au+Au collisions was observed by the STAR Collaboration at RHIC~\cite{STAR:2017ckg,Adam:2018ivw}. The data reveal that a QGP drop may possess a vorticity of the order of $10^{22}\,\text{s}^{-1}$, surpassing the vorticity of all other known fluids in nature~\cite{STAR:2017ckg}.

Besides the global $\Lambda$ polarization, another remarkable effect of the OAM is the {\it global spin alignment} of vector mesons~\cite{Liang:2004xn,Yang:2017sdk,Sheng:2019kmk,Sheng:2020ghv}. Following the idea of the global polarization, if quarks and anti-quarks in QGP are globally polarized along the OAM direction, vector mesons produced by quark recombination will have different probabilities to occupy spin states $S_y=1$, $0$, and $-1$. Here the $y$ axis is along the OAM direction, which is perpendicular to the reaction plane (the $z$-$x$ plane with $z$ axis along the colliding beams and $x$ axis along the impact parameter). In Ref.~\cite{Liang:2004xn}, it was found that the $00$-th element of the spin density matrix of the vector meson is related to the spin polarization of quarks and anti-quarks through:
\begin{equation}
    \rho_{00} = \frac{1-\Pqy\Pqbary}{3+\Pqy\Pqbary}.
    \label{rho00-old}
\end{equation}
In this equation, the $y$ axis has been chosen as the spin-quantization axis, and $\Pqy$ and $\Pqbary$ are the spin polarization of quarks and anti-quarks along the $y$ axis, respectively.

According to Eq.~(\ref{rho00-old}), if quarks and anti-quarks are globally polarized along the $y$ axis, $\rho_{00}$ will deviate from 1/3~\cite{Liang:2004xn,Yang:2017sdk,Sheng:2019kmk,Sheng:2020ghv}. Recently, the STAR and ALICE Collaborations reported the experimental results of $\rho_{00}$ for $\phi$ and $K^{*0}$ mesons which indeed deviates from $1/3$ in a wide range of centrality~\cite{Zhou:2019lun,Singha:2020qns,Acharya:2019vpe} but with an unexpectedly large magnitude [the centrality-averaged $|\Delta\rho_{00}|\sim\bigO(10^{-2})$ with $\Delta\rho_{00}=\rho_{00}-1/3$] that has not been understood. In fact, according to the global $\Lambda$ polarization data~\cite{STAR:2017ckg,Adam:2018ivw,Acharya:2019ryw}, the global polarization of quarks and anti-quarks is most likely to be only a few percent at $\sNN=\bigO(10)$ GeV and even smaller at $\bigO(10^3)$ GeV which, by Eq.~(\ref{rho00-old}), can only lead to a $|\Delta\rho_{00}|\sim\bigO(10^{-4})$ or smaller. This problem becomes more prominent in central collisions, in which the global polarization vanishes, but the STAR data still show a $\rho_{00}$ smaller than 1/3 in the most central events~\cite{Zhou:2019lun,Singha:2020qns}.

However, the above analysis based on the global polarization is not the entire story in realistic heavy-ion collisions, because the global OAM is not the only source of vorticity. In fact, the anisotropic expansion of the QGP can generate complicated local structure of the vorticity which does not contribute to the global OAM, see e.g.~\cite{Teryaev:2015gxa,Baznat:2015eca,Jiang:2016woz,Pang:2016igs,Li:2017slc,Becattini:2017gcx,Voloshin:2017kqp,Xia:2018tes,Wei:2018zfb,Ivanov:2017dff,Kolomeitsev:2018svb}, and then particles in QGP can be polarized locally, which leads to specific correlations between the particle polarization and momentum-space coordinates. This phenomenon for $\Lambda$ hyperon (called the {\it local $\Lambda$ polarization}) has been a subject of intense theoretical~\cite{Becattini:2017gcx,Voloshin:2017kqp,Xia:2018tes,Wei:2018zfb} and experimental studies~\cite{Adam:2019srw}, though the azimuthal-angle dependence of the local $\Lambda$ polarization remains an intriguing puzzle to be resolved~\cite{Becattini:2019ntv,Xia:2019fjf,Xie:2019jun,Florkowski:2019voj,Wu:2019eyi,Liu:2019krs}.

In this Letter, we propose a scenario of {\it local spin alignment} arising from the local vorticity structure. In this scenario, the spin alignment of vector mesons is caused by the local polarization of quarks and anti-quarks rather than their global polarization. We will show that, even in the situation of zero global polarization (e.g., in central collisions), the local polarization of quarks and anti-quarks can still drive $\rho_{00}$ of vector mesons to deviate from 1/3. We also investigate the characteristics of the local spin alignment and propose several observables that can distinguish between the local and the global spin alignments. We use $\hbar=c=1$.

\section{\label{sec-2}Spin density matrix of the vector meson}

The spin state of a vector meson can be described by a $3\times 3$ spin density matrix $\rho^V$:
\begin{equation}
    \rho^V = \begin{pmatrix}
        \rho_{11}  & \rho_{10}  & \rho_{1-1}  \\
        \rho_{01}  & \rho_{00}  & \rho_{0-1}  \\
        \rho_{-11} & \rho_{-10} & \rho_{-1-1}
    \end{pmatrix},
\end{equation}
where the indices $1$, $0$, and $-1$ label the spin component of the vector meson along the spin-quantization axis. To calculate $\rho^V$, we consider that vector mesons are formed by quark and anti-quark recombination, in which quarks and anti-quarks are polarized along arbitrary directions and their spin polarization vectors are
\begin{equation}
    \mathbf{P}^{q,\qbar} = (P_x^{q,\qbar}, P_y^{q,\qbar}, P_z^{q,\qbar}).
\end{equation}
Here all the components of $\mathbf{P}^{q,\qbar}$ can be nonzero. This is different from the scenario of global spin alignment in which only $P_y^{q,\qbar}$ is considered to be nonzero.

Next, we shall express the matrix elements of $\rho^V$ in terms of $P_x^{q,\qbar}$, $P_y^{q,\qbar}$, and $P_z^{q,\qbar}$. We first choose a specific direction as the spin-quantization axis. In principle, any direction can be chosen as the spin-quantization axis and the expression of $\rho^V$ would depend on this choice. In this Letter, we choose the $y$ axis as the spin-quantization axis unless noted otherwise. This choice is the same as the one in Eq.~(\ref{rho00-old}), and it allows us to study $\rho_{00}$ which is measured in experiment with respect to the $y$ axis (see explanation in Sec.~\ref{sec-4}).

Using the $y$ axis as spin-quantization axis, the spin density matrix of quarks and anti-quarks can be written as
\begin{equation}
    \rho^{q,\qbar}=\frac{1}{2}
    \begin{pmatrix}
        1+P_y^{q,\qbar}              & P_z^{q,\qbar}-iP_x^{q,\qbar} \\
        P_z^{q,\qbar}+iP_x^{q,\qbar} & 1-P_y^{q,\qbar}
    \end{pmatrix}.
    \label{rho-q}
\end{equation}
In the quark recombination mechanism, the produced vector meson is a spin-triplet state composed of its constituent quark and anti-quark. To obtain $\rho^V$, we first make a direct product of $\rho^q$ and $\rho^\qbar$ and then project it to the spin-triplet state~\cite{Liang:2004xn,Efremov:1981vs,Xu:2001hz}:
\begin{equation}
    \rho^V = \frac{U \rho^q \otimes \rho^\qbar U^\dagger}{\tr(U \rho^q \otimes \rho^\qbar U^\dagger)},
    \label{U-rho-U}
\end{equation}
where the transform matrix $U$ is
\begin{equation}
    U=\begin{pmatrix}
        1 & 0                  & 0                  & 0 \\
        0 & \frac{1}{\sqrt{2}} & \frac{1}{\sqrt{2}} & 0 \\
        0 & 0                  & 0                  & 1
    \end{pmatrix}.
\end{equation}
By inserting Eq.~(\ref{rho-q}) into Eq.~(\ref{U-rho-U}), we obtain the matrix elements of $\rho^V$:
\begin{align}
    \rho_{11}   & = \frac{(1+\Pqy)(1+\Pqbary)}{\ePP}, \label{rho11}                   \\
    \rho_{00}   & = \frac{1-\Pqy\Pqbary+\Pqx\Pqbarx+\Pqz\Pqbarz}{\ePP}, \label{rho00} \\
    \rho_{-1-1} & = \frac{(1-\Pqy)(1-\Pqbary)}{\ePP}, \label{rho-1-1}
\end{align}
and
\begin{align}
    \rho_{10}  & = \rho_{01}^*  = \frac{(1+\Pqy)(\Pqbarz-i\Pqbarx)+(\Pqz-i\Pqx)(1+\Pqbary)}{\sqrt{2}(\ePP)}, \label{rho10}  \\
    \rho_{0-1} & = \rho_{-10}^* = \frac{(1-\Pqy)(\Pqbarz-i\Pqbarx)+(\Pqz-i\Pqx)(1-\Pqbary)}{\sqrt{2}(\ePP)}, \label{rho0-1} \\
    \rho_{1-1} & = \rho_{-11}^* = \frac{(\Pqz-i\Pqx)(\Pqbarz-i\Pqbarx)}{\ePP}. \label{rho1-1}
\end{align}
From these equations, we observe that $\rho_{00}$ does not only receive contribution from the polarization component $P_y^{q,\qbar}$, but also from $P_x^{q,\qbar}$ and $P_z^{q,\qbar}$. Furthermore, in the presence of $P_x^{q,\qbar}$ and $P_z^{q,\qbar}$, the off-diagonal elements of $\rho^V$ can also be nonzero. However, these contributions from $P_x^{q,\qbar}$ and $P_z^{q,\qbar}$ were not taken into account in the previous studies on the spin alignment [cf., Eq.~(\ref{rho00-old})].

\section{\label{sec-3}Polarization vector of the vector meson}

From the spin density matrix $\rho^V$ obtained in Sec.~\ref{sec-2}, we can directly calculate the spin polarization of the vector meson by inserting Eqs.~(\ref{rho11}-\ref{rho1-1}) into $\mathbf{P}^V=\tr(\rho^V\,\widehat{\mathbf{S}})$, where the spin operators $\widehat{\mathbf{S}}$ for vector mesons are
\begin{equation}
    \widehat{S}_x = \frac{1}{\sqrt{2}}\begin{pmatrix}
        0 & -i & 0  \\
        i & 0  & -i \\
        0 & i  & 0
    \end{pmatrix},
    \ \widehat{S}_y = \begin{pmatrix}
        1 & 0 & 0  \\
        0 & 0 & 0  \\
        0 & 0 & -1
    \end{pmatrix},
    \ \widehat{S}_z = \frac{1}{\sqrt{2}}\begin{pmatrix}
        0 & 1 & 0 \\
        1 & 0 & 1 \\
        0 & 1 & 0
    \end{pmatrix}.
    \label{S_operators}
\end{equation}
Then we can obtain
\begin{equation}
    \mathbf{P}^V=\frac{2(\mathbf{P}^q+\mathbf{P}^\qbar)}{3+\mathbf{P}^q\cdot\mathbf{P}^\qbar}.
    \label{PV}
\end{equation}
This equation provides a simple relation between the spin polarization of the vector meson and the spin polarization of quarks and anti-quarks. It is interesting to compare Eq.~(\ref{PV}) with the hadron spin polarization obtained from the picture of the local thermodynamic equilibrium~\cite{Becattini:2013fla,Becattini:2013vja,Fang:2016vpj,Liu:2020flb}. In the latter picture, the spin polarization of a particle in a rotating fluid is determined by the fluid vorticity. For example, in non-relativistic limit, the polarization of a spin-$1/2$ particle is given by~\cite{Becattini:2016gvu}
\begin{equation}
    \mathbf{P}_{1/2}=\tanh\left(\frac{\omega}{2T}\right)\hat{\bm{\omega}},
    \label{P1/2}
\end{equation}
where $\omega$ is the strength of fluid vorticity $\bm\omega=(1/2)\bm\nabla\times\mathbf{v}$ with $\mathbf{v}$ the flow velocity, $\hat{\bm\omega}$ is the unit vector along the vorticity direction, and $T$ is the temperature. Then, with both the spin polarization of quarks and anti-quarks given by Eq.~(\ref{P1/2}), we obtain from Eq.~(\ref{PV})
\begin{equation}
    \mathbf{P}^V=\frac{2\sinh(\omega/T)}{1+2\cosh(\omega/T)}\hat{\bm\omega}.
    \label{pv-recom}
\end{equation}
This expression is exactly the spin polarization for a spin-1 particle obtained in Ref.~\cite{Becattini:2016gvu}.

Similarly, substituting Eq.~(\ref{P1/2}) for both $\mathbf{P}^q$ and $\mathbf{P}^{\bar{q}}$ in Eqs.~(\ref{rho11}-\ref{rho1-1}), we obtain
\begin{equation}
    \rho^V=\frac{1+\hat{\bm\omega}\cdot\widehat{\mathbf{S}}\sinh(\omega/T)+(\hat{\bm\omega}\cdot\widehat{\mathbf{S}})^2[\cosh(\omega/T)-1]}{1+2\cosh(\omega/T)},
    \label{dm-recom}
\end{equation}
which is equal exactly to
\begin{equation}
    \rho^V=\frac{1}{Z}e^{\bm\omega\cdot\widehat{\mathbf{S}}/T},
    \label{dm-direc}
\end{equation}
with $Z=\tr\exp{(\bm\omega\cdot\widehat{\mathbf{S}}/T)}$. The agreement between the expressions (\ref{pv-recom}) and (\ref{dm-recom}) from recombination formulas and the corresponding ones from direct calculations at local thermodynamic equilibrium is a result of angular momentum conservation during the process of recombination of quarks and anti-quarks into hadrons.

In experiments~\cite{Abelev:2008ag,Zhou:2019lun,Singha:2020qns,Acharya:2019vpe}, the spin information of $\phi$ and $K^{*0}$ mesons is extracted by analyzing their strong decays $\phi\to KK$ and $K^{*0}\to K\pi$. In these decays, the angular distributions of the decay products are even functions with respect to a reflection of $\mathbf{P}^V\to-\mathbf{P}^V$~\cite{Tang:2018qtu}, therefore the polarization vector $\mathbf{P}^V$ of $\phi$ and $K^{*0}$ meson is not measurable in the experiments. In fact, what can be measured in the spin density matrix are the variables listed in Eqs.~(\ref{rho00}, \ref{var-1}-\ref{var-4}), as we will discuss in Sec.~\ref{sec-4}.

\section{\label{sec-4}Measurable elements in the spin density matrix}

\begin{figure}
    \centering
    \includegraphics[width=1\columnwidth]{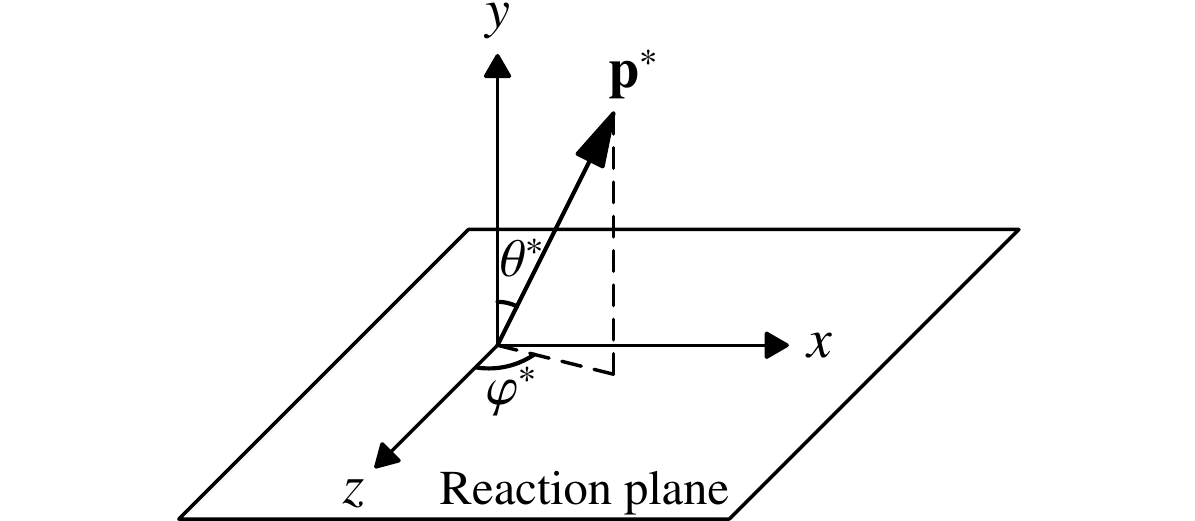}
    \caption{Definitions of $\theta^*$ and $\varphi^*$ in Eq.~(\ref{angular_dis}). Here $\mathbf{p}^*$ is the momentum of one decay product in the vector meson rest frame.}
    \label{fig-1}
\end{figure}

Now let us turn to the measurable elements in the spin density matrix $\rho^V$. As we mentioned in Sec.~\ref{sec-3}, the spin information of $\phi$ and $K^{*0}$ mesons is achieved by analyzing their decays $\phi\to KK$ and $K^{*0}\to K\pi$ in the experiments. In these decays, the angular distribution of the decay products is (see e.g., Ref.~\cite{Liang:2004xn})
\begin{align}
    \frac{d^2N}{d(\cos\theta^*)d\varphi^*}
    =\, & \frac{3}{8\pi}[(1-\rho_{00})+(3\rho_{00}-1)\cos^2\theta^*         \nonumber \\
        & -\sqrt{2}(\re\rho_{10}-\re\rho_{0-1})\sin(2\theta^*)\cos\varphi^* \nonumber \\
        & +\sqrt{2}(\im\rho_{10}-\im\rho_{0-1})\sin(2\theta^*)\sin\varphi^* \nonumber \\
        & -2\re\rho_{1-1}\sin^2\theta^*\cos(2\varphi^*)                     \nonumber \\
        & +2\im\rho_{1-1}\sin^2\theta^*\sin(2\varphi^*)],
    \label{angular_dis}
\end{align}
where $\theta^*$ and $\varphi^*$ are the polar and azimuthal angles of the momentum $\mathbf{p}^*$ of one decay product (see Fig.~\ref{fig-1} for the definition). The asterisk designates that $\mathbf{p}^*$, $\theta^*$, and $\varphi^*$ are defined in the vector-meson rest frame. Since we have chosen the $y$ axis as the spin-quantization axis when we derived $\rho^V$ in Sec.~\ref{sec-2}, the reference frame of $\theta^*$ and $\varphi^*$ has also been specified accordingly: $\theta^*$ is the angle between $\mathbf{p}^*$ and the $y$ axis, and $\varphi^*$ is the angle between the projection of $\mathbf{p}^*$ in the $z$-$x$ plane and the $z$ axis.

From Eq.~(\ref{angular_dis}), we see that there are five variables that can be determined in experiment by measuring the distributions of $\theta^*$ and $\varphi^*$. According to Eqs.~(\ref{rho11}-\ref{rho1-1}), these variables are related to $\mathbf{P}^q$ and $\mathbf{P}^\qbar$ by
\begin{align}
    -\sqrt{2}(\re\rho_{10}-\re\rho_{0-1}) & = -\frac{2(\Pqy\Pqbarz+\Pqz\Pqbary)}{\ePP}, \label{var-1} \\
    \sqrt{2}(\im\rho_{10}-\im\rho_{0-1})  & = -\frac{2(\Pqx\Pqbary+\Pqy\Pqbarx)}{\ePP}, \label{var-2} \\
    -2\re\rho_{1-1}                       & = -\frac{2(\Pqz\Pqbarz-\Pqx\Pqbarx)}{\ePP}, \label{var-3} \\
    2\im\rho_{1-1}                        & = -\frac{2(\Pqx\Pqbarz+\Pqz\Pqbarx)}{\ePP}, \label{var-4}
\end{align}
and $\rho_{00}$ has already been given in Eq.~(\ref{rho00}).

\section{Spin alignment in heavy-ion collisions}

In this section, we use the equations obtained in sections~\ref{sec-2} and~\ref{sec-4} to study the spin alignment of vector mesons in heavy-ion collisions. We discuss the global spin alignment and the local spin alignment separately because of their different sources of spin polarization of quarks and anti-quarks.

\subsection{Global spin alignment}

In noncentral collisions, it has been well known that the global OAM can cause the global polarization of quarks and anti-quarks along the $y$ axis. Then, through Eq.~(\ref{rho00-old}) or Eq.~(\ref{rho00}), one can obtain the global spin alignment $\rho_{00}$ as~\cite{Liang:2004xn}
\begin{equation}
    \rho_{00} = \frac{1-\PG^2}{3+\PG^2}\approx\frac{1}{3}-\frac{4}{9}\PG^2.
    \label{rho00-global}
\end{equation}
Here, we have assumed that quarks and anti-quarks have the same global polarization $\PG$, and the expression after ``$\approx$'' is obtained when $\PG$ is small.

From Eq.~(\ref{rho00-global}), we can observe that $\PG$ naturally causes $\rho_{00}$ to be less than 1/3. To estimate the value of $1/3-\rho_{00}$, we refer to the experimental data of the global $\Lambda$ polarization~\cite{STAR:2017ckg,Adam:2018ivw,Acharya:2019ryw}. According to the data, $\PG$ decreases with increasing beam energy: it changes from a few percent at $\sNN=$ 7.7 GeV to a few permille at 200 GeV, and it is consistent with zero at 2.76 and 5.02 TeV. Therefore, the value of $1/3-\rho_{00}$ contributed by $\PG$ should be of the order of $10^{-4}$ or less, which is not enough to explain the spin alignment of $\phi$ and $K^{*0}$ meson observed in experiments~\cite{Zhou:2019lun,Acharya:2019vpe,Singha:2020qns}.

Furthermore, because $\PG$ is caused by the global OAM, it only exists in noncentral collisions. Nevertheless, in the most central collisions, the experimental data~\cite{Zhou:2019lun,Singha:2020qns} still show evidence of $\rho_{00}\neq1/3$. Therefore, there must be other sources of the spin alignment in addition to the global OAM.

\subsection{Local spin alignment}

Besides the global OAM, the anisotropic expansion of QGP fireball is another source of the fluid vorticity, which leads to the local polarization. Below, we consider two specific kinds of the local polarization and give an intuitive interpretation of them. For more detailed discussion and analysis, we refer the reader to Refs.~\cite{Becattini:2017gcx,Voloshin:2017kqp,Xia:2018tes,Wei:2018zfb}.

First, in noncentral collisions, the expansion of the fireball has an ``elliptic flow'' on the transverse plane (the $x$-$y$ plane). This elliptic flow can generate the vorticity component in the $z$ direction by $\omega_z=(\partial_x v_y-\partial_y v_x)/2$, where $v_x$ and $v_y$ are the transverse velocity. Because the fireball at midrapidity $Y=0$ is symmetric under coordinate reflections $x\to-x$ and $y\to-y$, the longitudinal vorticity $\omega_z$ should present a quadrupole pattern on the transverse plane (see Refs.~\cite{Becattini:2017gcx,Voloshin:2017kqp} for details). In this longitudinal vorticity field, particles are polarized differently in each quadrant region. To the leading order of the harmonic series, this longitudinal local polarization can be expressed as~\cite{Becattini:2017gcx,Voloshin:2017kqp}:
\begin{equation}
    P_z(\Delta\psi) = F_z\sin(2\Delta\psi).
    \label{local-Pz}
\end{equation}
Here $\Delta\psi$ is the azimuthal angle of the particle's transverse momentum with respect to the reaction plane (the $x$-$z$ plane). Because this longitudinal local polarization is generated by the “elliptic flow”, the harmonic coefficient $F_z$ in Eq.~(\ref{local-Pz}) is nonzero only in noncentral collisions.

\begin{figure}
    \centering
    \includegraphics[width=0.7\columnwidth]{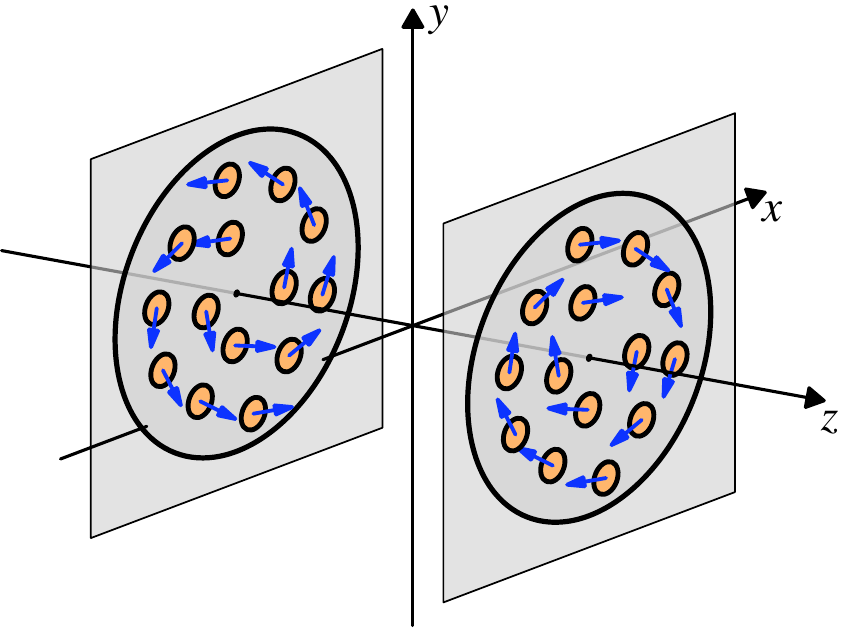}
    \caption{Illustration of the transverse local polarization in Eqs.~(\ref{local-Px}-\ref{local-Py}). The polarization vectors are along opposite orientations  (clockwise and counter-clockwise) at positive and negative longitudinal positions, respectively.}
    \label{fig-2}
\end{figure}

On the other hand, the transverse velocity $v_x$ and $v_y$ may also have gradients along the $z$ direction. Such gradient can be created from an initial non-uniform distribution of the fireball along the $z$ direction, and it can lead to the transverse vorticity components by $(\omega_x,\omega_y)=(-\partial_z v_y,\partial_z v_x)/2$. Because the transverse velocity in (non)central collisions is (approximately) along the radial direction in the cylindrical coordinate system, i.e.~$(v_x,v_y)=v_r\mathbf{e}_r$, it can be shown that the transverse vorticity is along the azimuthal direction, i.e.~$(\omega_x,\omega_y)=(\partial_z v_r)\mathbf{e}_\phi/2$ (see Refs.~\cite{Xia:2018tes,Wei:2018zfb} for details). This vorticity field forms a ``circular structure'' around the $z$ axis, in which particles get polarized along the vorticity loops, as illustrated in Fig.~\ref{fig-2}. To the leading order of the harmonic series, this transverse local polarization can be expressed as~\cite{Xia:2018tes,Wei:2018zfb}:
\begin{align}
    P_x(\Delta\psi) & = F_x\sin(\Delta\psi),  \label{local-Px} \\
    P_y(\Delta\psi) & = -F_y\cos(\Delta\psi). \label{local-Py}
\end{align}
Here the harmonic coefficients $F_x$ and $F_y$ are nonzero in both central and noncentral collisions. This is contrary to $\PG$ and $F_z$, which are nonzero only in noncentral collisions.

We can use the geometric symmetry of the heavy-ion collision system shown in Fig.~\ref{fig-2} to further constrain $F_x$ and $F_y$ in Eqs.~(\ref{local-Px}-\ref{local-Py}). First, due to a rotational symmetry around the $z$ axis, $F_x$ and $F_y$ should be equal to each other in central collisions; but in noncentral collisions, $F_x$ and $F_y$ are slightly different due to an elliptic geometry of the fireball~\cite{Xia:2018tes}. Besides, the system has also a rotational symmetry around the $y$ axis\footnote{In noncentral collisions, the $y$ axis is along the global OAM; while in central collisions, the $y$ axis can be defined as any direction perpendicular to the $z$ axis due to the rotational symmetry around $z$.} by an angle of 180 degrees in both central and noncentral collisions. Therefore, the polarization vectors in Fig.~\ref{fig-2} are along opposite orientations (clockwise and counter-clockwise) at positive and negative longitudinal positions, respectively. This implies that $F_x$ and $F_y$ are odd functions of rapidity $Y$. Thus, the transverse local polarization shown in Fig.~\ref{fig-2} vanishes at $Y=0$, but exists at finite rapidity.

Below, we investigate the effect of the local polarization on the spin alignment. We first consider the most central collisions, in which we have $\PG=F_z=0$ and $F_x=F_y\neq0$ as we discussed above, so the vector meson spin alignment in central collisions is caused by the transverse local polarization.

Then, to calculate the spin alignment $\rho_{00}$ in central collisions, we consider a vector meson at given $\Delta\psi$ at finite rapidity ($Y\neq0$). We assume that the vector meson is formed by recombination of a quark and an anti-quark which are at the same $\Delta\psi$ and $Y$. Then, by inserting Eqs.~(\ref{local-Px}-\ref{local-Py}) into Eq.~(\ref{rho00}), we obtain
\begin{align}
    \rho_{00}(\Delta\psi) & = \frac{1-F_\perp^2\cos(2\Delta\psi)}{3+F_\perp^2},                           \label{rho00-Fperp}  \\
                          & \approx \frac{1}{3}-\frac{F_\perp^2}{9}-\frac{F_\perp^2}{3}\cos(2\Delta\psi). \label{rho00-expand}
\end{align}
Here $F_\perp\equiv F_x=F_y$, and $\Delta\psi\equiv\psi_\text{vec}-\Psi_\text{RP}$ is the azimuthal angle of the vector meson with respect to the reaction plane. Equation (\ref{rho00-expand}) is obtained when $F_\perp$ is small. From Eqs.~(\ref{rho00-Fperp}-\ref{rho00-expand}), we observe that the transverse local polarization can lead to a deviation of $\rho_{00}$ from 1/3, and its value oscillates in $\Delta\psi$.

After integrating out $\Delta\psi$ in Eq.~(\ref{rho00-expand}), we obtain the average value of $\rho_{00}$:
\begin{equation}
    \langle\rho_{00}\rangle\equiv\frac{1}{2\pi}\int_0^{2\pi}\rho_{00}(\Delta\psi)d(\Delta\psi)\approx\frac{1}{3}-\frac{F_\perp^2}{9}.
    \label{rho00-mean}
\end{equation}
We see that the transverse local polarization can lead to $\langle\rho_{00}\rangle$ less than 1/3, although the polarization itself is zero after the average, i.e.~$\langle P_x\rangle=\langle P_y\rangle=0$ in Eqs.~(\ref{local-Px}-\ref{local-Py}).

Besides $\rho_{00}$, the transverse local polarization can also lead to nonzero off-diagonal elements of the spin density matrix $\rho^V$. By substituting Eqs.~(\ref{local-Px}-\ref{local-Py}) into Eqs.~(\ref{var-1}-\ref{var-4}), we find that the following two quantities are not vanishing:
\begin{align}
    \sqrt{2}(\im\rho_{10}-\im\rho_{0-1}) & = \frac{2F_\perp^2}{3+F_\perp^2}\sin(2\Delta\psi),  \label{local-var2} \\
    -2\re\rho_{1-1}                      & = \frac{2F_\perp^2}{3+F_\perp^2}\sin^2(\Delta\psi). \label{local-var3}
\end{align}

In Eqs.~(\ref{rho00-Fperp}-\ref{local-var3}), it is noteworthy that the contributions of $F_\perp$ are in terms of $F_\perp^2$. This quadratic power arises from the quark recombination in which a quark at $Y>0$ ($Y<0$) is assumed to combine with an anti-quark which is also at $Y>0$ ($Y<0$). As a result, although $F_\perp$ is an odd function of $Y$ according to Fig.~\ref{fig-2}, its contribution ($\propto F_\perp^2$) to Eqs.~(\ref{rho00-Fperp}-\ref{local-var3}) does not cancel after averaged over a finite rapidity range.

\begin{figure}
    \centering
    \includegraphics[width=0.9\linewidth]{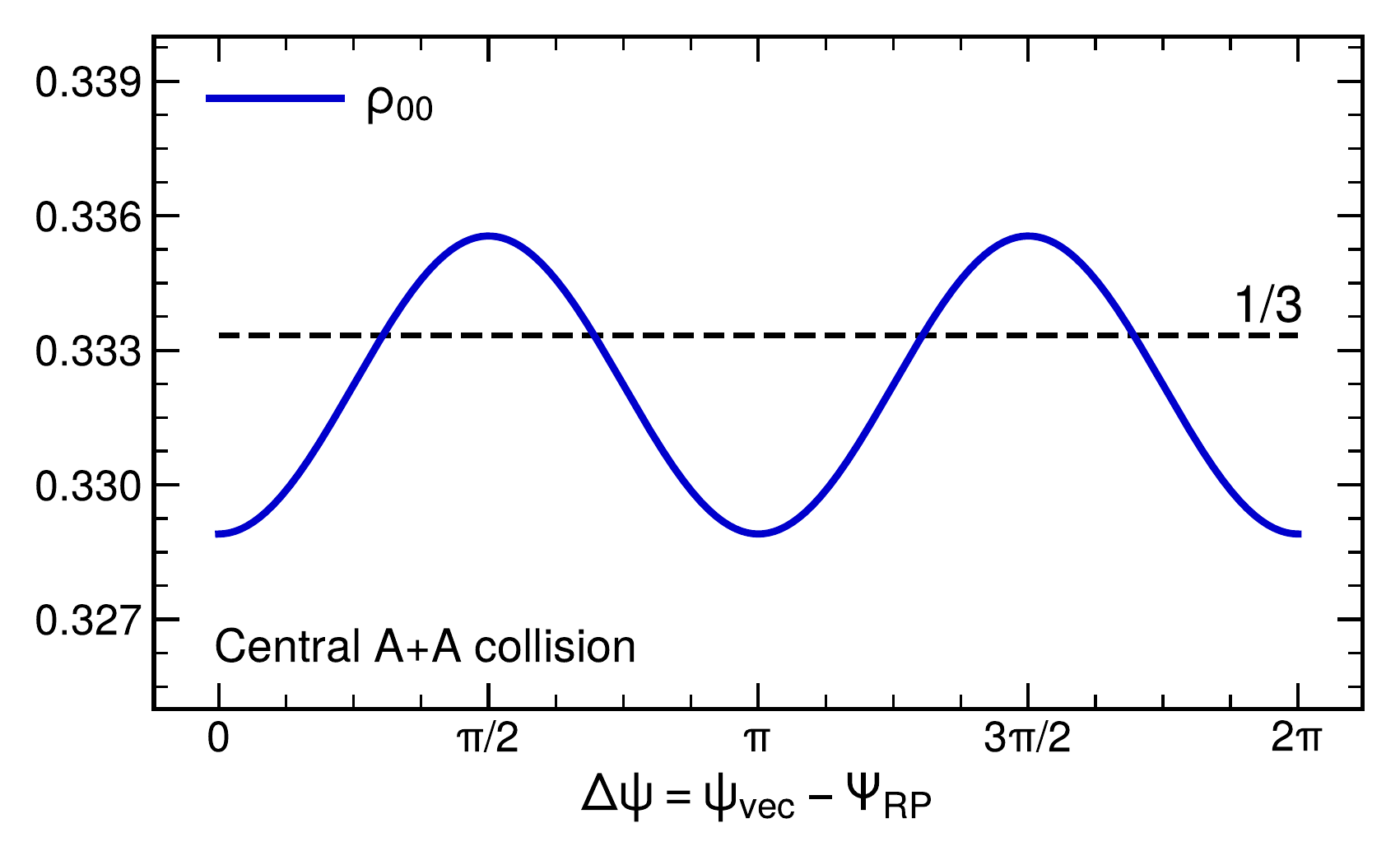}
    \includegraphics[width=0.9\linewidth]{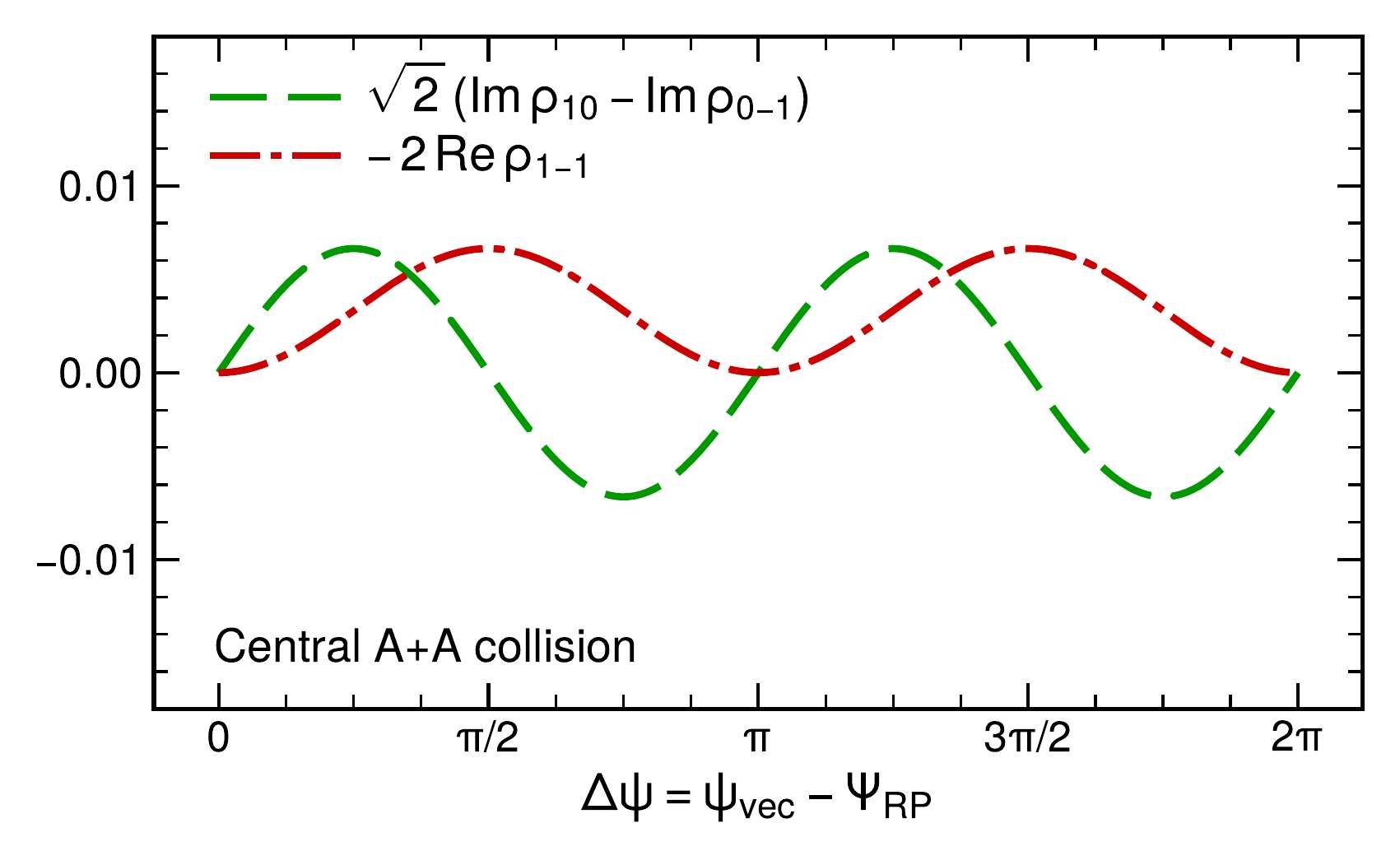}
    \caption{$\rho_{00}$ (upper panel), and $\sqrt{2}(\im\rho_{10}-\im\rho_{0-1})$ and $-2\re\rho_{1-1}$ (lower panel) as functions of $\Delta\psi$ in central collisions.}
    \label{fig-3}
\end{figure}

Fig.~\ref{fig-3} shows $\rho_{00}$, $\sqrt{2}(\im\rho_{10}-\im\rho_{0-1})$, and $-2\re\rho_{1-1}$ as functions of $\Delta\psi$. When we plot this figure, Eqs.~(\ref{rho00-Fperp}, \ref{local-var2}-\ref{local-var3}) are used and the value of $|F_\perp|$ is chosen to be 0.1. This value is estimated according to a previous theoretical calculation of the transverse local $\Lambda$ polarization in a finite rapidity range $|Y|<1$ in Au+Au collisions at 200 GeV, given in Ref.~\cite{Xia:2018tes}. Thus the results shown in Fig.~\ref{fig-3} should be considered as being averaged over rapidity region $|Y|<1$. We note that currently there is no experimental data yet to constrain $F_\perp$, and different theoretical models may give different $F_\perp$. However, the $\Delta\psi$ dependence of $\rho_{00}$, $\sqrt{2}(\im\rho_{10}-\im\rho_{0-1})$, and $-2\re\rho_{1-1}$ as shown in Fig.~\ref{fig-3} is a robust feature which is insensitive to the actual value of $F_\perp$.

For an intuitive understanding of the $\Delta\psi$ behavior of $\rho_{00}$ as shown in Fig.~\ref{fig-3}, let us recall the polarization pattern shown in Fig.~\ref{fig-2}, in which we see that particles at $\Delta\psi=0$ and $\pi$ are polarized along $\pm y$ directions, therefore, according to Eq.~(\ref{rho00}), $\rho_{00}$ at $\Delta\psi=0$ and $\pi$ is smaller than 1/3. On the other hand, particles at $\Delta\psi=\pi/2$ and $3\pi/2$ are polarized along $\pm x$ directions, thus the corresponding $\rho_{00}$ is larger than 1/3. By analyzing Eq.~(\ref{rho00}), one can also find that $\Pqx\Pqbarx+\Pqz\Pqbarz<2\Pqy\Pqbary$ is the condition for $\rho_{00}<1/3$. Therefore, solving this condition with the polarization pattern shown in Fig.~\ref{fig-2}, we find that the $\Delta\psi$ range for $\rho_{00}<1/3$ is wider than that for $\rho_{00}>1/3$, and thus $\langle\rho_{00}\rangle$ is smaller than 1/3 after taking the average over $\Delta\psi$.

Then, let us turn to noncentral collisions. In this case, $\PG$ and $F_z$ are nonzero, and $F_x$ and $F_y$ are unequal. Then, the vector meson spin alignment in noncentral collisions will receive contributions from both the global and the local polarization. By adding all the contributions together, we can write the total polarization of quarks and anti-quarks as
\begin{align}
    P_x(\Delta\psi) & = F_x\sin(\Delta\psi),     \label{noncentral-Px} \\
    P_y(\Delta\psi) & = \PG-F_y\cos(\Delta\psi), \label{noncentral-Py} \\
    P_z(\Delta\psi) & = F_z\sin(2\Delta\psi).    \label{noncentral-Pz}
\end{align}
Then, by inserting Eqs.~(\ref{noncentral-Px}-\ref{noncentral-Pz}) into Eqs.~(\ref{rho00}, \ref{var-1}-\ref{var-4}), one can calculate the elements of $\rho^V$ as functions of $\Delta\psi$. This would lead to more complicated $\Delta\psi$ modulations than those in central collisions. However, if the mean value of $F_x$ and $F_y$ can be much larger than $\PG$, $F_z$ and $F_x-F_y$, the $\Delta\psi$ dependence of the $\rho^V$ is still dominated by the one shown in Fig.~\ref{fig-3}.

In the rest of the section, we shall discuss the impact of strong magnetic field on our results. In the presence of the magnetic field, quarks and anti-quarks acquire additional spin polarization due to the coupling between their magnetic moments and the magnetic field. Hence, their spin polarization can be written as:
\begin{equation}
    \mathbf{P}^{q,\qbar} = \mathbf{P}_\omega^{q,\qbar} + \mathbf{P}_B^{q,\qbar}.
\end{equation}
Here, the vorticity induced polarization $\mathbf{P}_\omega^q$ and $\mathbf{P}_\omega^\qbar$ are along the same direction; but, for electrically neutral mesons such as $\phi$ and $K^{*0}$, the magnetic-field induced polarization $\mathbf{P}_B^q$ and $\mathbf{P}_B^\qbar$ are along opposite directions.

Constrained by the geometric symmetry of the collision system, the magnetic field created in heavy-ion collisions may have a similar spatial structure to the vorticity field. For instance, in noncentral collisions, there exists a global magnetic field that is mainly along the $y$ axis~\cite{Deng:2012pc,Bloczynski:2012en,Skokov:2009qp}, which can lead to the global polarization $\PG$. Besides, in both central and noncentral collisions, the expansion of the colliding system can produce a magnetic field circling the $z$ axis, which leads to a transverse polarization in a similar pattern of Fig.~\ref{fig-2}. Therefore, our analysis of the spin alignment caused by the vorticity can also be applied to the one caused by the magnetic field. The difference therein is that $\mathbf{P}_B^q$ and $\mathbf{P}_B^\qbar$ have the different signs. As a result, if the magnetic field dominates over the vorticity, the value of $\rho_{00}-1/3$ in Eqs.~(\ref{rho00-global}) and (\ref{rho00-Fperp}-\ref{rho00-mean}) would generally flip its sign, which makes $\langle\rho_{00}\rangle>1/3$.

\section{Global versus local spin alignments}

We have shown that local spin alignment as well as global spin alignment can all lead to deviation of $\langle\rho_{00}\rangle$ from 1/3. Below, we highlight the difference between the local and the global spin alignments, which can be used to distinguish these two scenarios in heavy-ion collisions.

(1) Although $\langle\rho_{00}\rangle$ deviates from 1/3 in both the local and the global spin alignments, the dependence of $\rho_{00}$ as a function of $\Delta\psi$ is different. In the local spin alignment, $\rho_{00}$ oscillates in $\Delta\psi$ between values larger than 1/3 and smaller than $1/3$; while in the global spin alignment, the sign of $\rho_{00}-1/3$ would be invariant versus $\Delta\psi$.

(2) In the local spin alignment, the quantities in Eqs.~(\ref{local-var2}-\ref{local-var3}) are nonzero. The amplitudes of their modulations ($\approx 2F_\perp^2/3$) are twice that of $\rho_{00}$ ($\approx F_\perp^2/3$). The existence of these nonzero off-diagonal elements of $\rho^V$ indicates that the angular distribution in Eq.~(\ref{angular_dis}) has a non-trivial shape not only in $\theta^*$ but also in $\varphi^*$. By contrast, the off-diagonal elements of $\rho^V$ are zero in the global spin alignment.

(3) In all the above discussions, we have taken the $y$ axis as the spin-quantization axis. Nevertheless, one can study how $\langle\rho_{00}\rangle$ changes with other choices of the spin-quantization axis. For example, we may choose the $x$ axis as the spin-quantization axis. Experimentally, this can be implemented by using the $y$-$z$ plane as the ``event plane'' and defining the $\theta^*$ angle with respect to the $x$ axis. In the local spin alignment, because the polarization pattern in Fig.~\ref{fig-2} has rotational symmetry around the $z$ axis, $\rho_{00}$ is independent from the choice of the event plane; see also the discussion in~\cite{Tang:2018qtu}. However, in the global spin alignment, $\langle\rho_{00}\rangle-1/3$ will flip its sign if the event plane is rotated from the $z$-$x$ plane to the $y$-$z$ plane.

\section{Summary}

We have studied the local spin alignment of vector mesons composed of quarks and anti-quarks which are locally polarized by the anisotropic expansion of the QGP fireball. We found that even if the global polarization of quarks and anti-quarks vanishes in the most central collisions, their local polarization can still cause $\rho_{00}$ of vector mesons to be smaller than 1/3, a feature that has been observed in experiments~\cite{Zhou:2019lun,Singha:2020qns}. Therefore, the observed $\rho_{00}\neq 1/3$ in the experiments does not establish unambiguously the global polarization of quarks and anti-quarks, but may also originate from the local polarization. We proposed that the measurements of $\Delta\psi$ dependence of $\rho_{00}$, off-diagonal elements in $\rho^V$, and $\langle\rho_{00}\rangle$ with respect to different event planes can be used to separate the local spin alignment from the global one in the experiments. The measurements of these observables will provide important information about the local vorticity structure of the QGP fireball and shed light on the puzzles in $\phi$ and $K^{*0}$ spin alignments.

\section*{Acknowledgments}

We thank Jinhui Chen, Chensheng Zhou, and Xin-Nian Wang for helpful discussions. This work is supported by National Natural Science Foundation of China through grants No.~11535012, No.~11675041, and No.~11835002. X.-L.~X.~and H.~L.~are also funded by China Postdoctoral Science Foundation through grants No.~2018M641909 and No.~2019M661333.

\bibliographystyle{elsarticle-num}
\bibliography{ref.bib}

\end{document}